\documentclass[fleqn,10pt]{wlscirep}
\usepackage[utf8]{inputenc}
\usepackage[T1]{fontenc}
\usepackage{subcaption}
\usepackage[capitalise]{cleveref}
\usepackage{multirow}
\usepackage{algorithm}
\usepackage{algpseudocode}

\algnewcommand\algorithmicinput{\textbf{Input:}}
\algnewcommand\algorithmicoutput{\textbf{Output:}}
\algnewcommand\Input{\item[\algorithmicinput]}%
\algnewcommand\Output{\item[\algorithmicoutput]}%
\title{VertXNet: An Ensemble Method for Vertebrae Segmentation and Identification of Spinal X-Ray}

\author[1,*]{Yao Chen}
\author[2,*]{Yuanhan Mo}
\author[1]{Aimee Readie}
\author[1]{Gregory Ligozio}
\author[3]{Indrajeet Mandal}
\author[3]{Faiz Jabbar}
\author[1,+]{Thibaud Coroller}
\author[2,+]{Bart\l omiej W. Papie\.z}
\affil[1]{Novartis Pharmaceuticals Company, East Hanover, NJ, USA}
\affil[2]{Big Data Institute, University of Oxford, Oxford, UK}
\affil[3]{John Radcliffe Hospital,Oxford University Hospitals NHS Foundation Trust, Oxford, UK}

\affil[+]{bartlomiej.papiez@bdi.ox.ac.uk, thibaud.coroller@novartis.com}

\affil[*]{these authors contributed equally to this work}
\affil[+]{these authors contributed equally to this work}

\begin{abstract}
Reliable vertebrae annotations are key to perform analysis of spinal X-ray images.
However, obtaining annotation of vertebrae from those images is usually carried out manually due to its complexity (i.e. small structures with varying shape), making it a costly and tedious process. 
To accelerate this process, we proposed an ensemble pipeline, VertXNet, that combines two state-of-the-art (SOTA) segmentation models (respectively U-Net and Mask R-CNN) to automatically segment and label vertebrae in X-ray spinal images.
Moreover, VertXNet introduces a rule-based approach that allows to robustly infer vertebrae labels (by locating the 'reference' vertebrae which are easier to segment than others) for a given spinal X-ray image. 
We evaluated the proposed pipeline on three spinal X-ray datasets (two internal and one publicly available), and compared against vertebrae annotated by radiologists.
Our experimental results have shown that the proposed pipeline outperformed two SOTA segmentation models on our test dataset (MEASURE 1) with a mean Dice of 0.90, vs. a mean Dice of 0.73 for Mask R-CNN and 0.72 for U-Net.
To further evaluate the generalization ability of VertXNet, the pre-trained pipeline was directly tested on two additional datasets (PREVENT and NHANES II) and consistent performance was observed with a mean Dice of 0.89 and 0.88, respectively.
Overall, VertXNet demonstrated significantly improved performance for vertebra segmentation and labeling for spinal X-ray imaging, and evaluation on both in-house clinical trial data and publicly available data further proved its generalization.

\end{abstract}

\begin{document}

\flushbottom
\maketitle
\thispagestyle{empty}
\section{Introduction}
X-ray imaging is a fast, non-invasive and low cost modality making it widely applied for spinal disease assessment and monitoring.
Reliable vertebrae annotation (i.e., vertebrae segmentation and identification) from spinal X-ray images is a prerequisite to quantitatively perform functional analysis of the spine, e.g., surgical planning or abnormality quantification\cite{burns2016automated}.
Over the past decade, deep learning has contributed greatly the field of computer vision, with many applications now become part of our everyday live, from facial recognition and indexing, photo stylizing or machine vision for self-driving cars\cite{lecun2015deep}. 
Many image segmentation and detection methods proposed in the last decade have been proven to be applicable for biomedical image analysis such as U-Net \cite{Ronneberger2015U-Net:Segmentation} and Mask R-CNN \cite{he2017mask}. 
However, there are still challenges on translating those tools to medical images.
As opposed to natural images, medical images (which includes spinal X-rays) have relatively limited spatial resolution. 
X-rays in particular are single plane images (compare to Magnetic Resonance Imaging (MRI), which is 3D imaging) that represent a patient's density projection and often exhibits a lower signal-to-noise ratio.
A unique challenge for vertebrae annotation in lateral spinal X-ray images is differentiating one vertebra from another vertebra as most would share similar shape and intensity (e.g. see the input images in \cref{fig:pipeline}). 
Data scarcity, commonly seen in medical applications, furthermore constrains the applicability of deep learning methods for the task, as a large amount of annotated training data is required for such methods to produce accurate results.
This is because manual annotation of X-ray images (e.g. labelling the landmarks of each vertebra) is non-trivial and must be performed by experts, which is time-consuming and expensive, but also prone to errors (i.e., inter- and intra-reader variability). 
Therefore, a solution that could automatically segment and identify the vertebrae in the spinal X-ray images would greatly decrease the cost, time, and inter-observer errors caused by human experts, and thus foster analysis of large X-ray imaging repositories coming e.g. from clinical studies.
\par
\subsection{Our Contribution}
To address these problems, we proposed a novel ensemble method VertXNet that combined two state-of-the-art (SOTA) segmentation and detection networks, namely \textbf{U-Net} and \textbf{Mask R-CNN}.
By combining the two networks, we can not only benefit from the advantages of semantic (U-Net) and instance (Mask R-CNN) segmentation networks in a single approach to produce robust segmentation results but also the identities of vertebrae can be accurately inferred in the final output according to the detection of the reference vertebra.
\par
The preliminary results of VertXNet were presented as a conference publication\cite{vx_miua} and this paper aims to extend the previous work as follows.
We provide detailed descriptions of our pipeline, including the key rule-based scheme that robustly combines segmentation outputs from both U-Net and Mask R-CNN. 
In order to demonstrate the generalizability, we tested our pipeline using datasets acquired from different centers and demonstrated that our pipeline is robust due to the ensemble architecture. 
Finally,  we also create extra labels (i.e., landmarks for the reference vertebrae) that are missing in the publicly available NHANES II dataset to enable the proposed pipeline to be compared to other methods.
To summarize:
\begin{enumerate}
\item In contrast to a conventional end-to-end deep-learning approach, our pipeline combined two SOTA networks (i.e. U-Net and Mask R-CNN) leveraging the advantages from both semantic and instance segmentation and showing a stronger performance than SOTAs alone.

\item Using a reference vertebra to infer the identities of the remaining vertebra according to their anatomical structure in X-ray images has shown greater performance and more robust prediction than predicting all vertebrae at once.

\item The segmentation and identification results of our pipeline was extremely robust i.e., the proposed pipeline has been tested on 3 different datasets including two in-house datasets from different clinical trials and one public dataset with the reference vertebrae annotated by our internal radiologists.

\item Reference vertebrae of a public dataset (NHANES II) were annotated by our internal radiologists.
The extra labels of reference vertebrae will be publicly shared alongside this investigation  in order to replicate our results and encourage other teams to advance the field of spinal segmentation and identification (The annotation for reference vertebrae of NHANES II is given in \cref{SecDA}). 
\end{enumerate}


\subsection{Related Work}
\label{SecRW}
Image segmentation is an attracting task thanks to the availability of larger datasets and its simplicity to debug (it is visually possible to debug our model performance by looking at the produced segmentation).
Before the booming development of deep learning, medical image segmentation methods were utilizing hand-crafted features and rule-based models for producing classification of each independent pixel \cite{masood2015survey}. 
\par
Traditional deformable models, e.g. Active Contour Models \cite{Kass1988Snakes:Models} (i.e., ACMs or so called 'Snakes'), deform the contours towards to the imaged object boundaries, which have proven to be effective methods.  
Zamora et al. \cite{zamora2003hierarchical} proposed to combine the generalized Hough transform, active shape models, and deformable models to address the segmentation task of cervical and lumbar vertebrae in spinal X-ray.
Xu et al. \cite{xu2007novel} proposed a method based on the Intersecting Cortical Model for segmenting the boundary of the spine from the frontal X-ray images.
Duong et al. \cite{duong2010automatic} proposed a support vector machine (SVM) based algorithm to segment the spinal region from posteroanterior (PA) X-ray images.
Fabian et al. \cite{lecron2012fully} developed a fully automatic approach for the vertebra detection, based on a SVM-model.
Mohamed Amine et al. \cite{larhmam2012semi} proposed an efficient modified template matching method for detecting cervical vertebrae using Generalized Hough Transform (GHT).
Ruhan et al. \cite{sa2016fast} proposed a novel framework that can accelerate a scale-invariant object detection method using a SVM which is trained on Histogram of Oriented Gradients (HOG), and then the segmentation of vertebra can be predicted using Gradient Vector Flow (GVF) based snake model.
Tezmol et al. \cite{tezmol2002customized} proposed a customized approach, based on the GHT, to produce the location and orientation of the cervical vertebrae boundaries in X-ray images.
\par
Recently, convolutional neural networks (CNNs) have become the most promising method for segmentation, from natural to medical images. \cite{de2015deep,ciresan2012deep,Ronneberger2015U-Net:Segmentation}.
Rashid et al. \cite{rashid2018fully} applied a fully convolutional neural network on segmenting lungs from chest X-rays. 
Novikov et al. \cite{novikov2018fully} also applied the convolutional neural networks for segmenting lungs, clavicles, and heart from chest X-rays. 
They demonstrated that the proposed pipeline was able to outperform the SOTA methods by combining a series of techniques such as delayed sub-sampling, exponential linear units and highly restrictive regularization.
Zhang et al. \cite{zhang2020mrln} proposed a multi-task relational learning network (MRLN) for vertebral localization, identification, and segmentation in Magnetic Resonance Images (MRI).
Pang et al. \cite{pang2020spineparsenet} proposed a two-stage framework, named SpineParseNet, for automatic spine parsing for MR images, where a 3D graph CNN was used for coarse segmentation and a 2D residual U-Net was then used for refining the segmentation.
Winsor et al. \cite{windsor2020convolutional} proposed a novel approach for labelling vertebrae sequences based on language-modelling. 
The proposed approach reached SOTA performance over a range of clinical datasets.
\par
However, the existing work for CNN based vertebral segmentation on X-ray images is still limited because most of the current CNN-based methods take MRI images as inputs \cite{nie2018asdnet}. 
There are still many challenges for fully automatic vertebrae segmentation and identification in spinal X-ray images.
Arif et al. \cite{al2018spnet} developed a novel network to predict shapes instead of performing pixel-wise classification for vertebra segmentation in X-ray images.
Tran et al. \cite{tran2020mbnet} proposed the MBNet which was a multi-task deep neural network for semantic segmentation on lumbar vertebrae.
Li et al. \cite{li2016automatic} proposed a novel CNN model that combined two types of features of lumbar vertebrae X-ray images to automatically detect lumbar vertebrae for C-arm X-ray images.
Kurachka et al. \cite{kurachka2017vertebrae}  proposed to use a CNN based method trained on patches from spinal X-ray images for vertebrae detection.
The model produced the likelihood of a vertebra being contained by the given patch.
Li et al. \cite{li2022spa} proposed a novel and computation-efficient network, called SPA-ResUNet, which combined residual U-Net with strip-pooling attention mechanism for the multi-class vertebrae and inter-vertebral discs segmentation.
Whitehead et al. \cite{whitehead2018deep} proposed a coarse-to-fine pipeline for spinal segmentation which used a series of four pixel-wise segmentation networks to refine the segmentation results step by step.
Bidur et al. \cite{cobbdetect} proposed an automatic deep learning-based method that firstly detects the vertebrae and then predicts 4 landmark corners for each vertebra to finally produce the Cobb Angles for spinal X-ray images.
Ruhan et al. \cite{sa2017intervertebral} fine tuned a pre-trained faster R-CNN on small annotated X-ray dataset, and they demonstrated that the faster R-CNN can outperform the traditional sliding window detection method with hand-crafted features.
\par

\section{Methodology}
\label{sec:method}

\begin{figure}[!h]
  \includegraphics[width=1\linewidth]{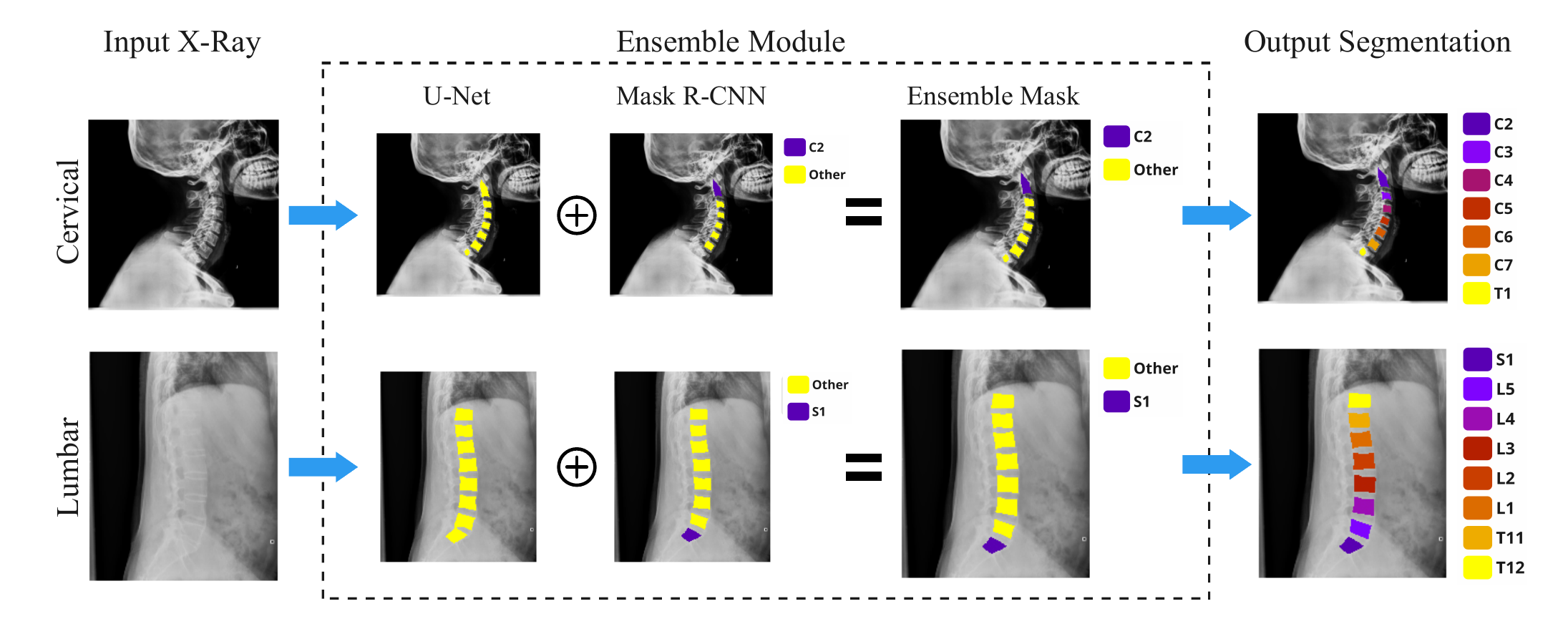}
  \caption{VertXNet overview: 1) An ensemble method (U-Net and Mask R-CNN) produces a robust segmentation for each vertebra, 2) Mask R-CNN locates a reference vertebra (either C2 or S1) and infers remaining vertebrae.}
  \label{fig:pipeline}
\end{figure}
The proposed pipeline, VertXNet, is trained and tested on either anonymized X-ray images (MEASURE 1 and PREVENT) from completed and anonymized secukinumab clinical trials or a public dataset (NHANES II).
As shown in \cref{fig:pipeline}, VertXNet employed a two-steps pipeline, \textit{segmentation} and \textit{identification} of spinal vertebrae.
X-ray images were fed into a segmentation module to first generate a separable segmentation mask of each vertebra.
Two SOTA segmentation networks, U-Net \cite{he2017mask} and Mask R-CNN \cite{he2017mask}, have been investigated.
However, according to our preliminary results, neither of them worked perfectly individually.
Thus, the two networks are trained independently in order to produce their own predictions. 
Then, a rule-based ensemble method is introduced to combine outputs from both networks to robustly produce both vertebrae segmentation and identification. 
From there, we inferred the identifications of the rest vertebrae starting from the detected reference vertebrae (one per image).

\subsection{U-Net}
Semantic segmentation aims to produce labels for each pixel over an image with a corresponding class of what is being represented.
Unlike instance segmentation, semantic segmentation is usually not able to distinguish different objects within one class.
\par
In the last decade, many deep learning-based networks for image segmentation have been proposed, which were mentioned in Section~\ref{SecRW}.
However, U-Net is the most representative one among these networks and has been widely used in biomedical image analysis.
U-Net is based on a fully convolutional network and its architecture was modified and extended to work with fewer training images and to yield more precise segmentation.
It consists of a contracting path and an expansive path.
\par
\begin{figure}
     \raggedright
     \begin{subfigure}[b]{0.45\textwidth}
         \centering
    \includegraphics[width=1.1\linewidth]{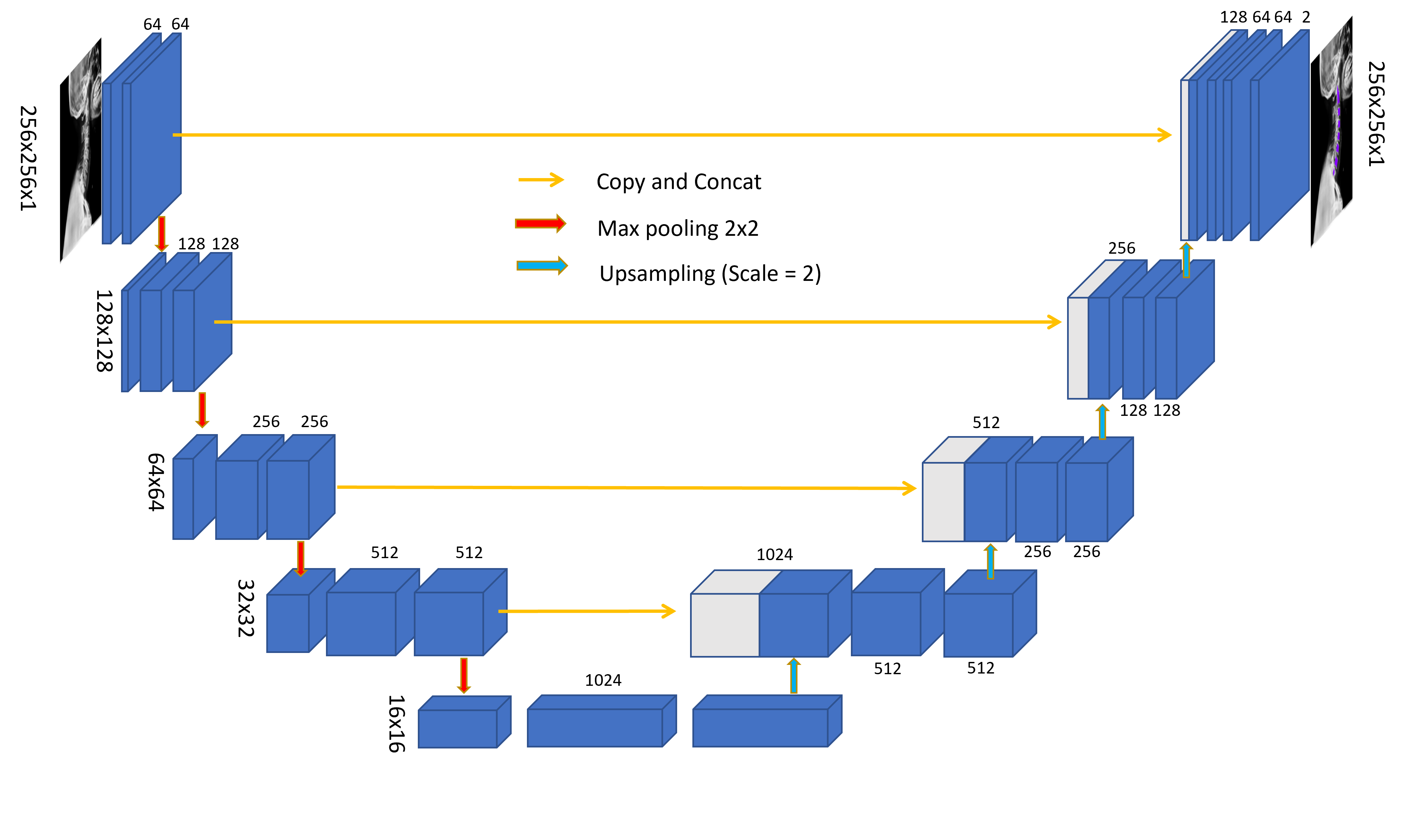}
            \caption{The U-Net utilized in our pipeline.}
  \label{FigUnet}
     \end{subfigure}
     \begin{subfigure}[b]{0.45\textwidth}
         \raggedleft
            \includegraphics[width=1.1\linewidth]{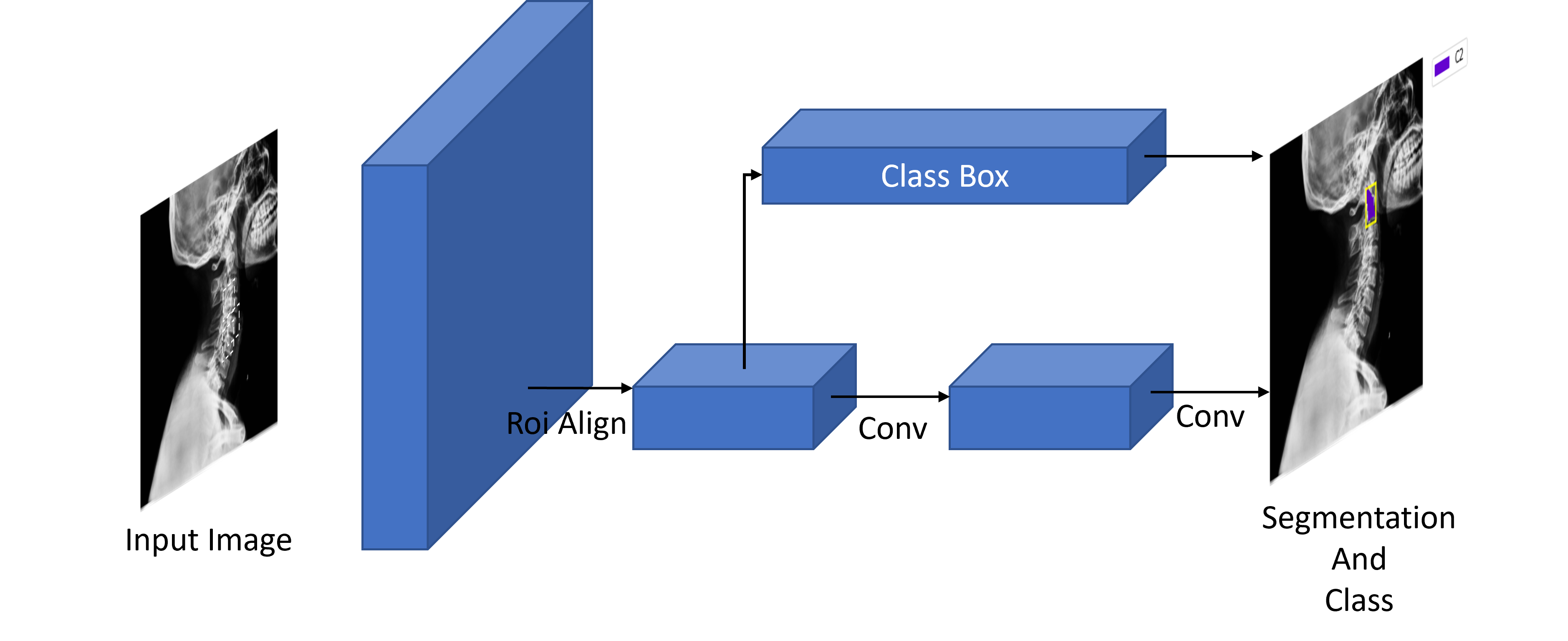}
            \caption{The pipeline of Mask R-CNN.}
            \label{fig:maskrcnn}
     \end{subfigure}
     \caption{Two segmentation models used in our pipeline, namely U-Net and Mask R-CNN.}
     \label{fig:SOTA}
\end{figure}

Our model use a modified version of the original U-Net as illustrated in \cref{FigUnet}.

The downscale path has 5 convolutional blocks, each composed of two convolutional layers with a filter size of 3×3, stride of 1, padding of 1 in both directions and followed by batch normalization and ReLU (Rectified Linear Unit) activation and batch normalization. 
Max pooling with kernel size of 2 is applied at the end of each block.
At the bottleneck, the number of feature maps increases from 1 to 1024.
In the upscale path, every block starts with a upsample layer with scale factor of 2, which doubles the size of feature maps in both directions but decreases the number of feature maps by two.
Two convolutional layers of filter size 3x3 and padding of 1 are applied on the concatenation of upsample feature maps and the feature maps from encoding path followed by batch normalization and ReLU activation. 
Max pooling of kernel size 2 is used at the end of each block.
The last layer is a 1×1 convolutional layer to collapse the number of features and predict our foreground / background classes (binary segmentation).

\subsection{Mask R-CNN}
As introduced in the previous section, semantic segmentation aims to categorize each pixel in an image into a fixed set of class without differentiating each individual object instances. Thus, it only addresses the identification/classification of same objects as a single class at the pixel level.
In contrast, instance segmentation addresses the issue of correct detection of all objects in an image, at the same time accurately segmenting each instance within one class.
Therefore, it is a series of operations including object detection, object localization, object segmentation and object classification. 
\par
R-CNN is considered as the first two-stage detector developed based on the convolutional neural network by proposing the Regions with CNN features (RCNN) for object detection\cite{obj20}.
Since then many CNN-based object detectors start to emerges at an unprecedented speed including fast R-CNN \cite{fastRCNN}, faster R-CNN \cite{fasterRCNN} and Mask R-CNN \cite{he2017mask}.
\par
In our pipeline, Mask R-CNN does not only detect the objects in an image, but also  simultaneously generates a precise segmentation mask for each instance (i.e., vertebrae).
Mask R-CNN is an extension of faster R-CNN \cite{fasterRCNN} which has two predictions for each candidate object, a category and an offset for bounding box.
Mask R-CNN adds a third branch which produces the object mask as an additional output besides the bounding box and its category.
The additional output of segmentation mask is different from the category and bounding box outputs, requiring the extraction of a much finer spatial layout of an object. 
\par
In our pipeline, we used the pre-trained components provided in PyTorch library \cite{pytorchpaper} for developing the Mask R-CNN.
A pre-trained ResNet-50 with Feature Pyramid Network (FPN) is used as the backbone of our Mask R-CNN.


\subsection{Ensemble Rules}
\begin{figure}
     \centering
     \begin{subfigure}[b]{0.45\textwidth}
         \centering
         \includegraphics[width=\textwidth]{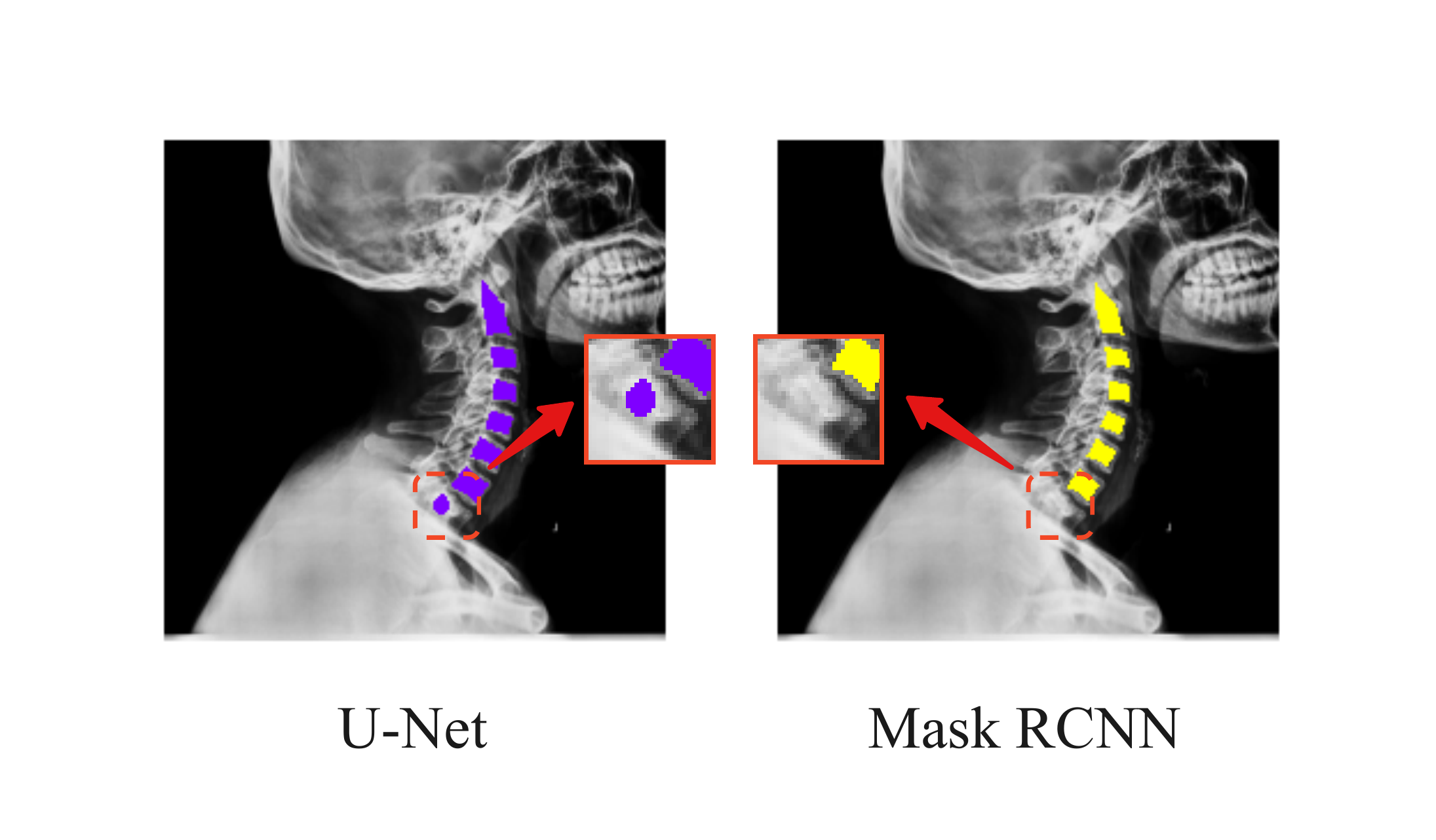}
         \caption{Missing Vertebra Example}
         \label{fig:mv}
     \end{subfigure}
     \begin{subfigure}[b]{0.45\textwidth}
         \centering
         \includegraphics[width=\textwidth]{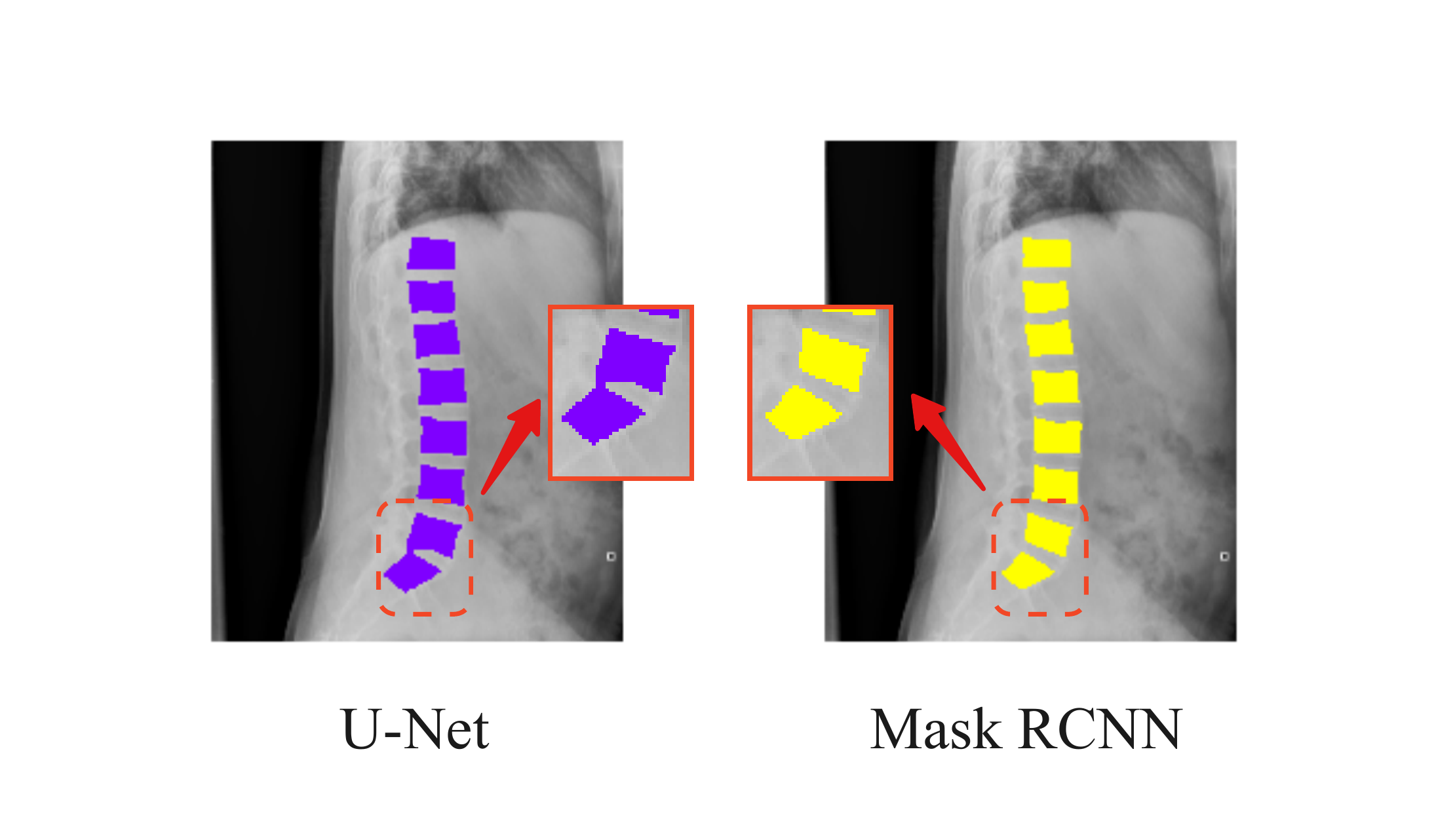}
         \caption{Overlapping Masks Example}
         \label{fig:om}
     \end{subfigure}
     \caption{Failure cases with individual SOTA method; (a) Mask R-CNN fails to detect vertebra T1 due to the incomplete of shape, however, U-Net successfully annotates the partial shape of vertebra T1; (b) U-Net generates overlapping contours and cannot be easily separated, however, Mask R-CNN generates mask within bounding box which makes all vertebrae separable.}
     \label{fig:SOTA_failure}
\end{figure}

\begin{figure}[!htbp]
\centering
  \includegraphics[width=0.75\linewidth]{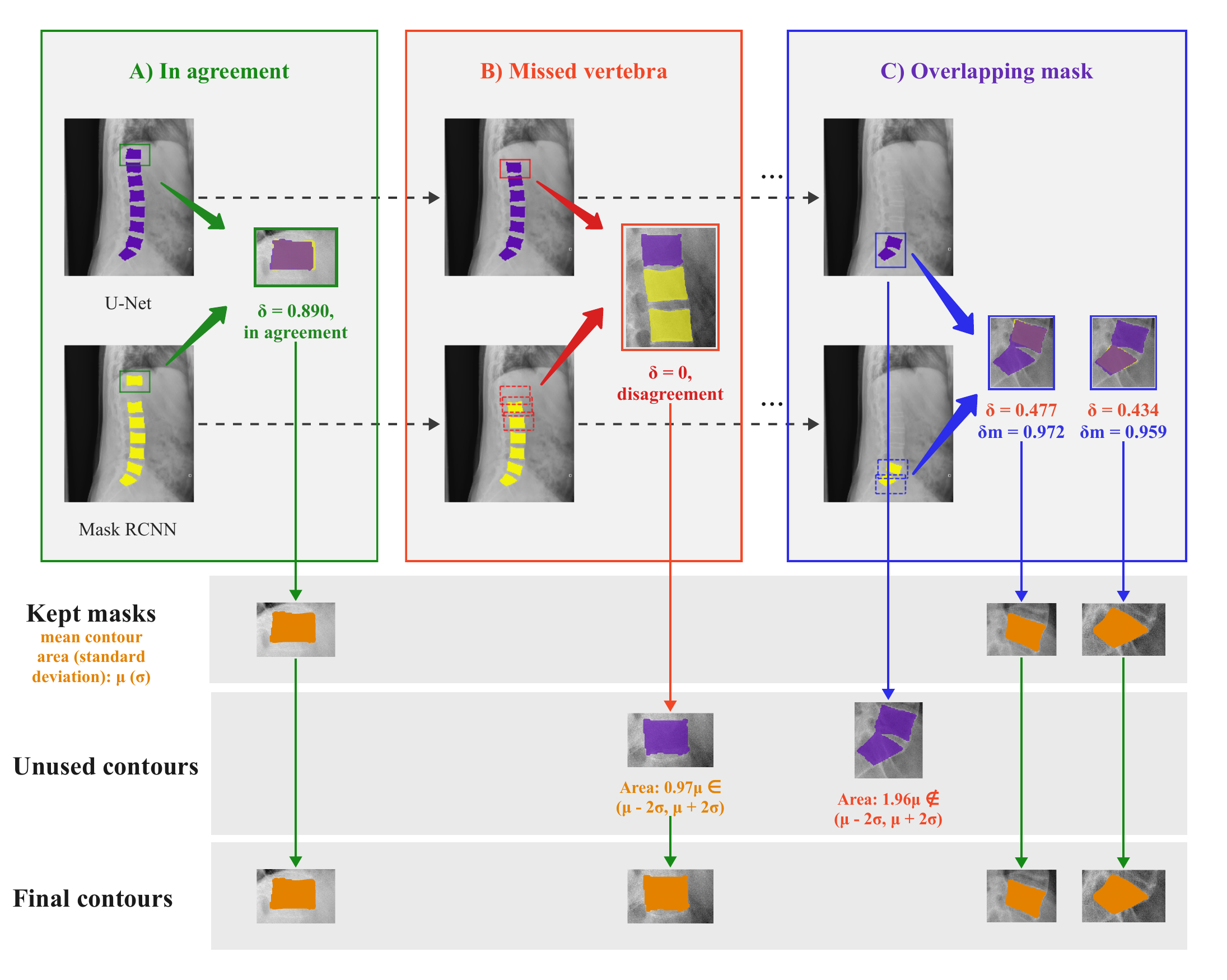}
  \caption{Ensemble procedure: Pairwise comparison between two segmentation results are performed; \textbf{A)} if two masks are in agreement, the union of two masks will be added to the list of ensemble masks, and the original masks will be removed from each segmentation; \textbf{B)} if the mask cannot find the mask in agreement/partial agreement from the other segmentation, it will be added to the remaining contours for further check; \textbf{C)} if Mask R-CNN segmentation overlapped by a large proportion with U-net but not the other way around, the Mask R-CNN generated mask will be added to the list of ensemble masks. At the end of each procedure, the mean and standard deviation of ensemble masks will be calculated and valid masks in remaining contours with similar size will be added to the ensemble masks list.}
  \label{fig:ensemble}
\end{figure}

In our preliminary experiments, neither of these two independent methods (U-Net and Mask R-CNN) could perfectly solve the vertebrae segmentation task (\cref{fig:SOTA_failure}). Previous works \cite{meng2022vertebrae, koo2022pilot} also showed the limitation of the single methods and required pre-process of the medical image, modification of framework or incorporation of prior knowledge.
We noticed that Mask R-CNN may have missed the prediction for those vertebrae, which have the incomplete boundary since the first step of object detection fails to identify the bounding box of the incomplete vertebrae,
while U-Net is more reliable to find vertebrae as it performs binary classification at pixel level (\cref{fig:mv}). 
However, under the setting of binary classification, U-Net may generate overlapping masks when two vertebrae in the X-ray image are located closely to each other. 
Moreover, unlike Mask R-CNN performing segmentation within bounding box, it is challenging for U-Net to separate the overlapping masks as illustrated in \cref{fig:om}.
Therefore, the proposed pipeline would jointly consider the outputs from both U-Net and Mask R-CNN in order to address the challenges faced by each individual model (see \cref{algo:ensemble} and \cref{fig:ensemble}).

{\bf Selected candidate masks in agreement.}
In the ensemble method, we first obtain output masks from both U-Net and Mask R-CNN, and extract the mask contours from each output. U-Net contours are denoted as $\{C^{u}_1, ..., C^{u}_m\}$ and Mask R-CNN contours $\{C^{m}_1, ..., C^{m}_n\}$. Then we define agreement between two contours by 
$$\delta(C_i, C_j) = \dfrac{area(C_i \cap C_j)}{max(area(C_i), area(C_j))} > \eta$$
$\eta$ is a pre-defined threshold value, and we recommend 0.6 as the cutoff. The choice can be flexible within a range, in our experiment, $\eta$ from 0.3 to 0.7 generated similar results. Then a cross comparison between contours of U-Net and contours of Mask R-CNN, i.e. $\{(C^{u}_i, C^{m}_j)|$ for all $i \in 1, ..., n$ and $j \in 1, ..., m\}$, are performed to select the candidate vertebral masks. If the a mask contour from U-Net and Mask R-CNN is selected as in agreement, the union of these pair of contours will be used in the final list of vertebral masks. 

{\bf Separate overlapping masks}. After the first step, not all candidate masks have been selected. Neither of the missing masks and overlapping masks from the scenario we demonstrated in \cref{fig:SOTA_failure} will be selected as candidate masks. We first targeted overlapping masks. We checked again the intersection of two contours given by U-Net and Mask R-CNN. If the intersection occupies a good percentage, e.g., $\eta$ defined in the first step, of Mask R-CNN contours, but relatively small on U-Net contours, it is very likely that this is an overlapping mask on U-Net. Here we denote $\delta_m(C^u_i, C^m_j) = \dfrac{area(C^u_i \cap C^m_j)}{area(C^m_j)}$ and use criteria $\delta < \eta$ and $\delta_m > \eta$ to define the finding of overlapping mask. In such case, we can directly take contours from Mask R-CNN as the candidate masks.

{\bf Pick up missing vertebrae.} Finally, we have one more challenge to solve, missing vertebrae by one method, usually from Mask R-CNN. In order to pick up missed vertebrae from remaining plausible contours, we need to use the selected contours in agreement for further check. The area of the candidate masks will be measured and the mean $\mu_C$ and standard deviation $\sigma_C$ of the list will be calculated. The more possible contours will be selected based on the area of the contours that are within $\lambda \cdot \sigma_C$ range of $\mu_C$. The $\lambda$ is a hyper-parameter to be selected. 

{\bf Sort all candidate masks.} After all three steps, we now have a complete set of vertebral masks that have worked around the issues faced by either individual method. 
Then the masks are sorted vertically on the image for the next step of inferring and labelling.

\begin{algorithm}
    \caption{Pseudo code for vertebrae ensemble rules}
    \begin{algorithmic}
        \Require Trained U-Net, Mask R-CNN
        \Input Spinal X-ray
        \Output list of ensemble contours $C^e$
        \State Pass X-ray to U-Net and Mask R-CNN, and extract contours $C^u = \{C^{u}_1, ..., C^{u}_m\}$ and $C^m = \{C^{m}_1, ..., C^{m}_n\}$
        \For{i in 1:m}
            \For {j in 1:n}
                \If{$\delta(C^u_i, C^m_j) = \dfrac{area(C^u_i \cap C^m_j)}{max(area(C^u_i), area(C^m_j))} > \eta$}
                    \State Add $C^u_i \cup C^m_j$ to $C^e$
                    \State Remove $C^u_i$ and $C^m_j$ from $C^u$ and $C^m$
                \ElsIf{$\delta_m(C^u_i, C^m_j) = \dfrac{area(C^u_i \cap C^m_j)}{area(C^m_j)} > \eta$}
                    \State Add $C^m_j$ to $C^e$   
                    \State Remove $C^m_j$ from and $C^m$
                \EndIf
            \EndFor
        \EndFor
        \State Calculate $\mu_C = mean(C^e)$, $\sigma_C = sd(C^e)$
        \While{$C^u$ or $C^m$ is not empty}
            \If{$area(C_i) \in (\mu_C - \lambda \sigma_C, \mu_C + \lambda \sigma_C)$}
                \State Add $C_i$ to $C^e$
            \EndIf
            \State Remove $C_i$
        \EndWhile
    \end{algorithmic}
\label{algo:ensemble}
\end{algorithm}

\subsection{Vertebrae Identification}
Due to the similarity in shape, it is challenging to correctly label the vertebrae under multi-class task settings (see Table \ref{tab:metrics}). The spatial relationship among vertebrae, such as the knowledge that L1 (lumbar \#1 vertebra) should be followed by L2 (lumbar \#2 vertebra), can be used. Thus finding the reference vertebrae becomes essential to properly label all remaining vertebrae in an X-ray image. 
Luckily, both lumbar and cervical images contain two types of vertebrae that can be easily distinguished from others, and are identified as the reference vertebrae.
The reference vertebrae include: 1) ‘C2’, cone-shaped and the first detectable vertebra at the top of the cervical X-ray image, and 2) ‘S1’, a triangular-shaped vertebra, which is commonly the last visible vertebra at the bottom of a lumbar X-ray. To detect the reference vertebrae, the Mask R-CNN is trained to distinguish the reference vertebra from other vertebrae.
If C2 is detected on an image, we simply "zip" down the spine to infer the cervical (C) and thoracic (T) labels (from C3 to C7 and T1). 
On the other hand, if "S1" is detected we "zip" up the spinal vertebrae to infer the lumbar (L) and thoracic (T) labels (from L5 to L1, and T12 to T11).

\section{Experiments and Results}
\label{SecEandR}

\subsection{Datasets} 

The proposed method was developed and evaluated on three spinal X-ray image datasets: two anonymized secukinumab axSpA clinical trials (MEASURE 1 and PREVENT) and a public dataset (NHANES II).

\subsubsection*{MEASURE 1}
A sample of images from the anonymized MEASURE~1 study \cite{baeten2015secukinumab} were utilized, in which sagittal cervical and lumbar X-ray images were acquired at different visits (Baseline, Week 104, and Week 208).
All vertebrae in the 512 X-ray images (293 cervical and 219 lumbar) were annotated by our internal radiologists. 


\begin{figure}
    \centering
    \begin{subfigure}[b]{0.45\textwidth}
         \centering
         \includegraphics[width=\textwidth]{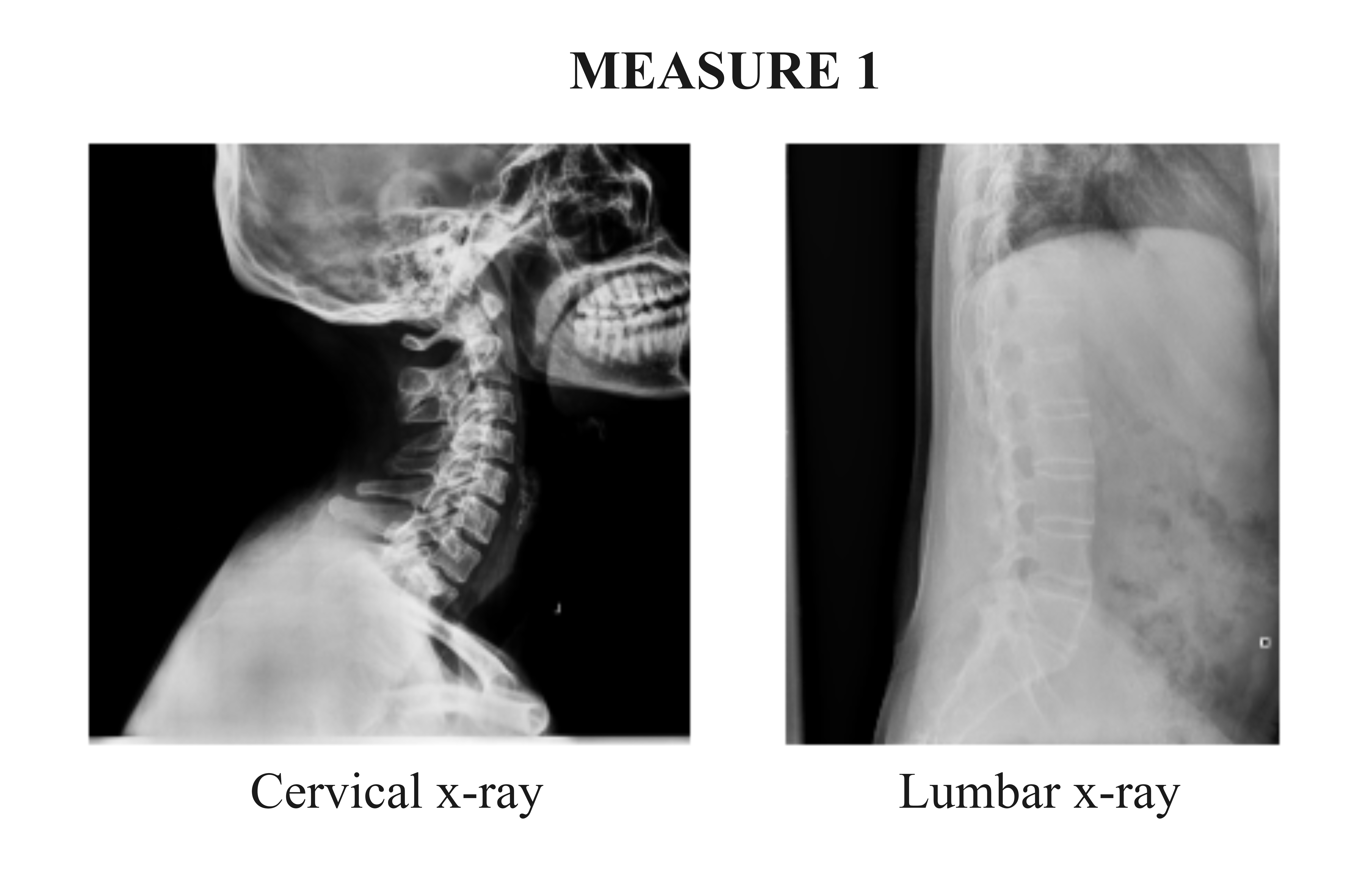}
         \caption{Examples of MEAUSRE 1 X-ray}
         \label{fig:measure}
     \end{subfigure}
     \begin{subfigure}[b]{0.45\textwidth}
         \centering
         \includegraphics[width=\textwidth]{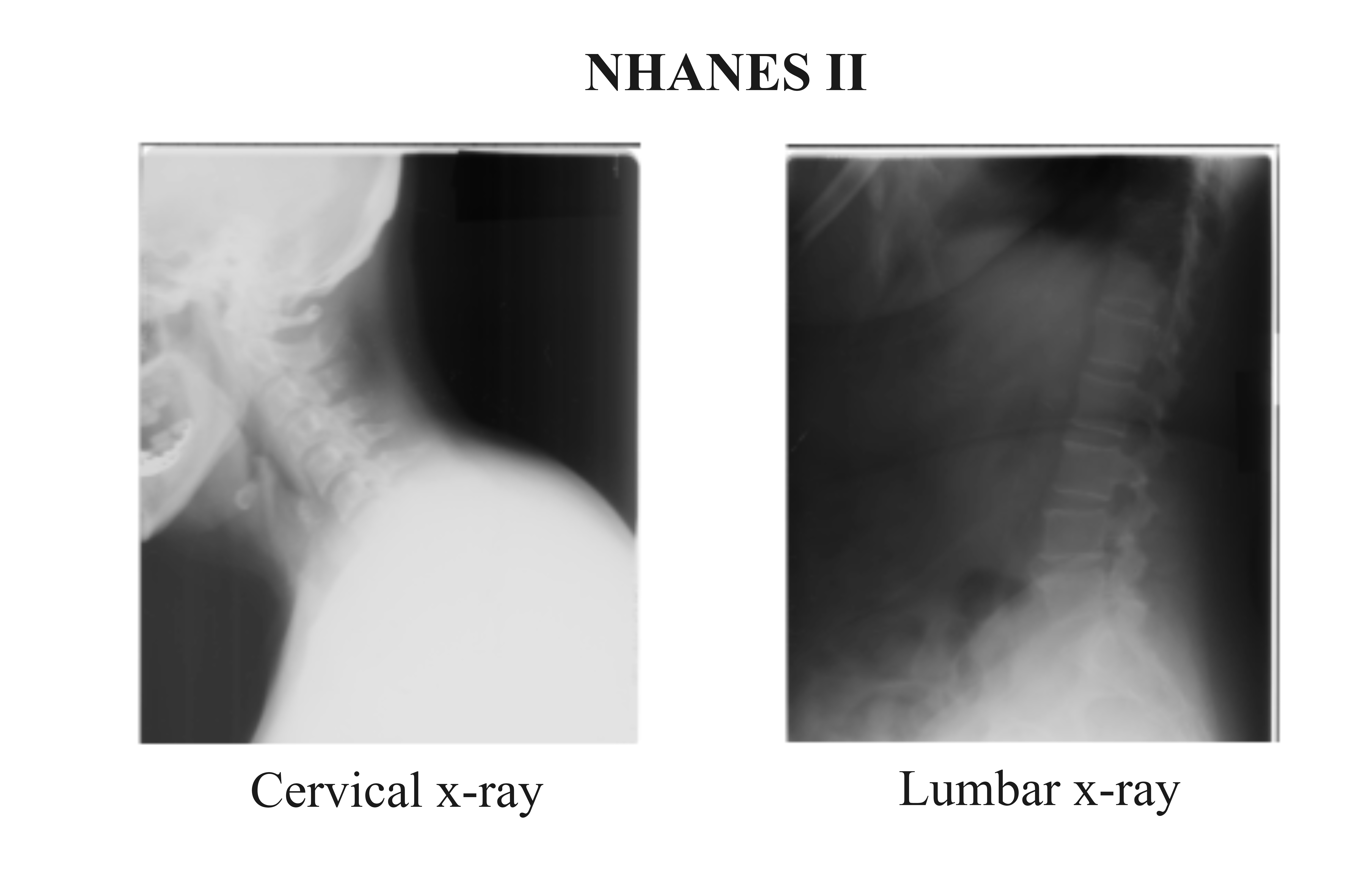}
         \caption{Examples of NHANES II X-ray}
         \label{fig:NHANES}
     \end{subfigure}
     \caption{Examples of X-rays from clinical trial and public dataset; MEASURE 1 X-rays have better quality in terms of resolution, contrast, etc. in comparing with NHANES II. Quality of X-rays are similar in PREVENT and MEASURE 1.}
     \label{fig:example}
\end{figure}


\subsubsection*{PREVENT}
Another anonymized in-house clinical trial dataset with a different population of axSpA patients, PREVENT  \cite{deodhar2021improvement} was also investigated. 
Visible vertebrae from 226 lateral X-rays (132 cervical and 94 lumbar) were annotated by the same internal radiologists.
Compared with MEASURE~1, the X-ray films in PREVENT captured slightly wider field of view of the spinal anatomy, which meant more T1, T11 and T12 were visible in the PREVENT dataset.

\subsubsection*{NHANES II}
\label{sec:nhanes}
During the Second National Health and Nutrition Examination Surveys (NHANES II), 17,000 lateral cervical/lumbar X-ray films were collected to provide evidence of osteoarthritis and degenerative disc disease. 
The NHANES II dataset also provided landmark annotation for vertebrae in 544 images, but only C3-C7 and L1-L5 were provided with complete landmark coordinates to generate masks. 
Our internal radiologists annotated the reference vertebrae (i.e., C2 and S1) when they were visible in the dataset. 
Eventually, we got 445 out of 544 images with reference vertebrae. 
The quality of the X-ray film in NHANES II is not as high as our in-house clinical datasets.


\subsection{Experiments}

\hspace{\parindent} {\bf Ensemble vs Individual Methods} The 512 annotated X-rays from MEASURE 1 were randomly split and stratified by spinal acquisition type.
80\% of the X-rays images were used for training the pipeline and the remaining 20\% for testing.
All models were trained on the training dataset and the Dice coefficients were calculated on the testing set to compare all models. 
Our pipeline's mean Dice coefficient was {\bf 0.90}, compared with Mask R-CNN's {\bf 0.73} and U-Net's {\bf 0.72}. Detailed Dice coefficients on each vertebral type have been provided in Table \ref{tab:metrics}. 
A drop in performances across models were observed for T1, T11 and T12 due to data imbalance caused by the lack of appearance of these vertebrae in X-rays. 
Greater performance from our model was observed for T1, T11 and T12.
In the experiment, the ensemble mask not only solved the potential risk of overlapping masks or missing vertebral masks by one model, but also improved the Dice coefficients of the outputs.

\begin{table}[!htbp]
    \begin{subtable}[h]{0.9\textwidth}
        \centering
        \begin{tabular}{lccccccc}
            \hline
            Model & C2 & C3 & C4 & C5 & C6 & C7 & T1 \\
               \hline\hline
             Mask R-CNN & 0.852 & 0.882 & 0.889 & 0.842 & 0.767 & 0.736 & 0.325 \\
            \hline
            U-Net & 0.823 & 0.846 & 0.848 & 0.822 & 0.809 & 0.717 & 0.067 \\
            \hline
            VertXNet & \bf 0.905 & \bf 0.912 & \bf 0.912 & \bf 0.906 & \bf 0.901 & \bf 0.881 & \bf 0.518 \\
            \hline
        \end{tabular}
        \caption{Dice coefficients of Cervical vertebrae on test set}
    \end{subtable}
    \begin{subtable}[h]{0.9\textwidth}
        \centering
        \begin{tabular}{lcccccccc}
            \hline
            Model & T11 & T12 & L1 & L2 & L3 & L4 & L5 & S1 \\
            \hline\hline
            Mask R-CNN & 0.379 & 0.577 & 0.706 & 0.730 & 0.615 & 0.707 & 0.732 & 0.744 \\
            \hline
            U-Net & 0.338 & 0.519 & 0.643 & 0.726 & 0.785 & 0.772 & 0.726 & 0.603 \\
            \hline
            VertXNet & \bf 0.761 & \bf 0.843 & \bf 0.864 & \bf 0.870 & \bf 0.871 & \bf 0.865 & \bf 0.854 & \bf 0.823 \\
        \end{tabular}
        \caption{Dice coefficients of Lumbar vertebrae on test set}
    \end{subtable}
    \caption{Model Comparison (numbers in bold indicate the best performance); The performance on the test set between VertXNet and the two benchmark models was demonstrated in this table. VertXNet surpassed both benchmark models in all vertebrae categories.}
    \label{tab:metrics}
\end{table}

\begin{figure}
     \centering
     \begin{subfigure}[b]{0.35\textwidth}
         \centering
         \includegraphics[width=\textwidth]{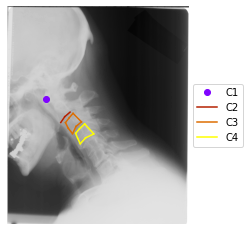}
         \caption{NHANES II annotation}
     \end{subfigure}
     \begin{subfigure}[b]{0.36\textwidth}
         \centering
         \includegraphics[width=\textwidth]{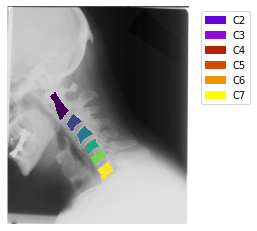}
         \caption{VertXNet prediction}
     \end{subfigure}
     \caption{One of the reasons that the evaluation metrics (i.e., Dice coefficients and Panoptic quality score) dropped on the NHANES II dataset was due to some incomplete annotations on NHANES II that have made the evaluation unfair when they were treated as ground truth. In Figure 6(a) shows an example of NHANES II annotation with only lower edge of C2, C3, C4, but in Figure 6(b), VertXNet has correctly segmented and labelled C2-C7.}
     \label{fig:incorrect_label}
\end{figure}

{\bf Generalizability of pre-trained pipeline} The pre-trained pipeline on the MEASURE~1 dataset was also tested on PREVENT and NHANES II to evaluate the performance. 
The Dice coefficients are reported in \cref{tab:generalization}. 
The overall Dice coefficients are similar among the two datasets, with {\bf 0.89} on PREVENT and {\bf 0.88} on NHANES II. 
A boost of performance on T1 of PREVENT was maybe seen due to the wider visibility range on the films, similarly for T11. NHANES II experienced some drop in performance mainly due to two reasons, very different quality (see \cref{fig:example}) and missing annotation on some vertebrae (see \cref{fig:incorrect_label}).

\begin{table}[!htbp]
    \begin{subtable}[h]{0.9\textwidth}
        \centering
        \begin{tabular}{lccccccc}
            \hline\hline
            Dataset & C2 & C3 & C4 & C5 & C6 & C7 & T1 \\
            \hline 
            MEASURE 1 & \bf 0.905 & \bf 0.912 & \bf 0.912 & \bf 0.906 & \bf 0.901 & \bf 0.881 & 0.518 \\
            \hline
            PREVENT & 0.900 & 0.904 & 0.899 & 0.875 & 0.863 & 0.873 & \bf 0.828 \\
            \hline
            NHANES II & 0.860 & 0.855 & 0.845 & 0.827 & 0.784 & 0.613 & -- \\
            \hline
        \end{tabular}
        \caption{Dice coefficients of Cervical vertebrae on test set}
    \end{subtable}
    \begin{subtable}[h]{0.9\textwidth}
        \centering
        \begin{tabular}{lcccccccc}
            \hline
            Dataset & T11 & T12 & L1 & L2 & L3 & L4 & L5 & S1 \\
            \hline\hline
            MEASURE 1 & 0.761 & \bf 0.843 & \bf 0.864 & \bf 0.870 & \bf 0.871 & \bf 0.865 & 0.854 & 0.823 \\
            \hline
            PREVENT & \bf 0.764 & 0.824 & 0.840 & 0.844 & 0.850 & 0.849 & \bf 0.893 & 0.847 \\
            \hline
            NHANES II & -- & -- & 0.802 & 0.821 & 0.820 & 0.823 & 0.865 & \bf 0.849 \\
            \hline
        \end{tabular}
        \caption{Dice coefficients of Lumbar vertebrae on test set}
    \end{subtable}
    \caption{Generalization of pre-trained pipeline on other datasets (numbers in bold indicate the best performance); The pre-trained pipeline on MEASURE 1 has been tested on PREVENT and NHANES II to show the generalization capability of VertXNet. The pre-trained pipeline has demonstrated robust results on datasets with different patient populations (PREVENT) or different image quality (NHANES II) when comparing with MEASURE 1.}
    \label{tab:generalization}
\end{table}

In some cases, for example in \cref{fig:incorrect_label}, no annotation from C3-C7 was provided in NHANES II, but the pipeline correctly detected those vertebrae. 
However, numerically the Dice coefficients dropped to 0 when compared with their annotation. 
The panoptic quality, which was raised in \cite{Kirillov_2019_CVPR}, is also provided for the comparison in \cref{tab:panoptic_quality}.

\begin{table}[!htbp]
    \begin{subtable}[h]{0.9\textwidth}
        \centering
        \begin{tabular}{lcccccccc}
            \hline
            Dataset & thresholding value & C2 & C3 & C4 & C5 & C6 & C7 & T1 \\
            \hline \hline
            \multirow{2}{*}{MEASURE 1} & 0.8 & 0.682 & 0.851 & 0.810 & 0.826 & 0.795 & 0.776 & 0.396 \\
            & 0.7 & 0.830 & 0.877 & 0.866 & 0.855 & 0.853 & 0.826 & 0.396 \\
            \hline
            \multirow{2}{*}{PREVENT} & 0.8 & 0.757 & 0.828 & 0.839 & 0.821 & 0.809 & 0.826 & 0.717 \\
            & 0.7 & 0.834 & 0.872 & 0.873 & 0.848 & 0.857 & 0.861 & 0.717 \\
            \hline
            \multirow{2}{*}{NHANES II} & 0.8 & 0.387 & 0.748 & 0.793 & 0.722 & 0.720 & 0.668 & -- \\
            & 0.7 & 0.673 & 0.851 & 0.844 & 0.808 & 0.789 & 0.706 & -- \\
            \hline
        \end{tabular}
        \caption{Panoptic Quality of Cervical vertebrae on test set}
    \end{subtable}
    \begin{subtable}[h]{0.9\textwidth}
        \centering
        \begin{tabular}{lccccccccc}
            \hline
            Dataset & thresholding value & T11 & T12 & L1 & L2 & L3 & L4 & L5 & S1 \\
            \hline\hline
            \multirow{2}{*}{MEASURE 1} & 0.8 & 0.702 & 0.788 & 0.888 & 0.896 & 0.912 & 0.828 & 0.789 & 0.654 \\
            & 0.7 & 0.762 & 0.837 & 0.888 & 0.896 & 0.912 & 0.900 & 0.860 & 0.767 \\
            \hline
            \multirow{2}{*}{PREVENT} & 0.8 & 0.778 & 0.832 & 0.866 & 0.897 & 0.913 & 0.913 & 0.831 & 0.624 \\
            & 0.7 & 0.855 & 0.864 & 0.890 & 0.897 & 0.913 & 0.913 & 0.975 & 0.786 \\
            \hline
            \multirow{2}{*}{NHANES II} & 0.8 & -- & -- & 0.727 & 0.796 & 0.797 & 0.816 & 0.793 & 0.501 \\
            & 0.7 & -- & -- & 0.833 & 0.873 & 0.862 & 0.881 & 0.843 & 0.745 \\
            \hline
        \end{tabular}
        \caption{Panoptic Quality of Lumbar vertebrae on test set}
    \end{subtable}
    \caption{Panoptic Quality; Panoptic quality was utilized to further demonstrate the robustness of the performance of VertXNet on all three datasets. The performance on both MEASURE 1 and PREVENT was good and similar with each other, performance on NHANES II dropped slightly due to the low quality of the images and ground truth.}
    \label{tab:panoptic_quality}
\end{table}

\section{Discussion}
\label{SecDiss}
Annotating vertebrae from spinal X-ray images is a crucial step to further investigate clinical diagnosis prediction (i.e. disease status, treatment response).
Despite a large number of studies of medical image segmentation in the literature \cite{masood2015survey,jamaludin2017spinenet}, there are still aforementioned challenges in developing an automatic and robust pipeline with high performance using limited training samples. 
\par
\begin{figure}[!h]
  \includegraphics[width=0.95\linewidth]{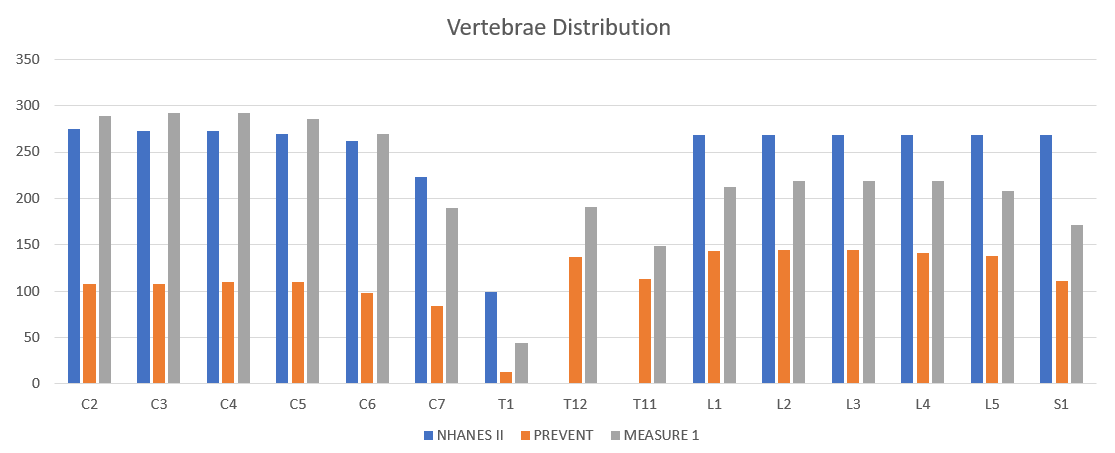}
  \caption{Vertebrae distribution of three datasets: NHANES II, PREVENT and MEASURE 1}
  \label{fig:vud}
\end{figure}
In this paper, we introduced an ensemble pipeline (VertXNet) to automatically segment and annotate vertebrae from spinal X-rays (working on either lumbar or cervical).
VertXNet combines two SOTA segmentation models (i.e., U-Net and Mask R-CNN) using a rule-based method, and demonstrated strong performances in our test sets. 
We evaluated our pipeline on three different datasets, namely MEASURE1, PREVENT and NHANES II, where two of them are from a secukinumab internal clinical trial at Novartis and the latter is public available with our own annotations to fit the experiments.
We demonstrated that the ensemble method significantly outperformed the U-Net and Mask R-CNN respectively using an unseen portion of our internal datasets.
Great and robust performance was also observed on the different internal datasets and public dataset with different patient population using pre-trained pipeline.
\par
Despite the promising results of our pipeline, there are still some limitations:
1) our pipeline used U-Net and Mask R-CNN to produce segmentation masks for vertebrae independently. 
Our ensemble rule was then used to merge the segmentation masks to produce the final results.
The ensemble rule was manually crafted and could have potential issue to not generalize well to other datasets or simply underperform.
Thus, it is worthy exploring that replacing the manual designed ensemble rule with a data-driven method. 
For example, we can introduce an extra neural network which implicitly learns the ensemble rule and merge the segmentation results.
This would allow us to avoid manually designing the ensemble rule and make the pipeline more generalized. 
2) Currently the U-Net and Mask R-CNN are trained respectively and independently. 
This is time consuming and inefficient compared to an end-to-end approach of training the models together. 
Training both models together would reduce the computation time while enabling us to dynamically merge their output, as mentioned in the previous point. 
This would however require further hyperparameter tuning and model designing, beyond the scope of this proof of concept for spinal segmentation.
3) Our proposed pipeline performed well to identify the names for each vertebra. However, this was achieved by locating the a reference vertebra (i.e., S1 and C2) first and then inferring the rest vertebrae according to their relative position with respect to the reference vertebra. This means that our pipeline breaks if we are unable to either detect a reference vertebra or we predicted the wrong location of the reference vertebra. 
During the experiments, our pipeline will fail to work on some cases when the reference vertebra are not detected in X-ray images (e.g. a X-ray scan does not cover the S1 or C2).
4) Some of the vertebrae like T1 and T12 are less frequently seen in the X-ray scans compared to others. 
This may provide our pipeline with fewer samples of these vertebrae during the training and the pipeline would underperform on segmenting them.
\cref{fig:vud} demonstrates the vertebrae distribution over three datasets.
We can see that the number of vertebrae like T1 and T12 is significantly less than other vertebrae which might be a reason that the measures (i.e., Dice and Panoptic Quality) decrease for T1, T11 and T12.
\par
In this study, we proposed an ensemble pipeline that can automatically segment and label vertebrae from lateral spinal X-ray images.
The pipeline combines two SOTA segmentation models, namely U-Net and Mask R-CNN, to produce robust segmentation mask and their corresponding identities.
Such a combination allows us to achieve promising vertebrae segmentation results compared to an individual SOTA model.
Our pipeline was evaluated on three spinal X-ray datasets, namely MEASURE 1, PREVENT and NHANES II, with the annotations provided by our internal experts.
Two measures (i.e., Dice score and Panoptic Quality) were used for the evaluation.
We demonstrated that our pipeline has outperformed the benchmark models on MEASURE 1 and can generalized well on PREVENT and NHANES II.
In the future, we aim to simplify the proposed pipeline and make the training process end-to-end.
Currently, the proposed pipeline required a reference vertebra to infer the names of other vertebrae. 
We will also look to make our pipeline less dependent on reference vertebra detection such that the pipeline can directly predict the names of each vertebra.
\section{Data Availability}
\label{SecDA}
Two of the datasets (MEASURE 1 and PREVENT) used in this study were collected from completed, anonymized clinical trials. 
Since the X-ray images were anonymized, these are no longer personal data and so no further EC/IRB (Ethics Committees/Institutional Review Board) approvals are required.
\par
Another used dataset (NHANES II) is a public dataset that was conducted by the National Center for Health Statistics, Centers for Disease Control (NCHS/CDC) during the Second National Health and Nutrition Examination Survey.
Further information about this dataset is available at \url{https://www.nlm.nih.gov/databases/download/nhanes.html}.
The extra labels of reference vertebrae for NHANES II will be available at \url{...} in the camera-ready version.
\section{Author Contribution}
Yao Chen and Yuanhan Mo implemented the relevant experiments and wrote the main manuscript. 
Indrajeet Mandal and Faiz Jabbar annotated the reference vertebrae of X-ray images in the public dataset.
All authors reviewed the manuscript and provided their valuable feedback.
\section{Additional Information}
The authors declare no competing interests.
\bibliography{sample}

\end{document}